\documentclass[nature,
                preprint,
                amsmath,
                amssymb,
                showpacs,
                superscriptaddress]{revtex4-1}
\usepackage{graphicx} 
\usepackage{dcolumn}  
\usepackage{bm}       
\usepackage{amsfonts}
\usepackage{dsfont}

\usepackage{ulem}
\usepackage{color}

\begin{document}

\title{Light-Induced Valleytronics  in Pristine Graphene}

\author{M. S. Mrudul}
\affiliation{%
Department of Physics, Indian Institute of Technology Bombay,
            Powai, Mumbai 400076, India }
            
\author{\'Alvaro Jim\'enez-Gal\'an}
\affiliation{%
Max-Born Institut, Max-Born Stra{\ss}e 2A, 12489 Berlin, Germany}

\author{Misha Ivanov}
\affiliation{%
Max-Born Institut, Max-Born Stra{\ss}e 2A, 12489 Berlin, Germany}
                        
\author{Gopal Dixit}
\email[]{gdixit@phy.iitb.ac.in}
\affiliation{%
Department of Physics, Indian Institute of Technology Bombay,
            Powai, Mumbai 400076, India }

\date{\today}


\begin{abstract}
Electrons in two-dimensional hexagonal materials have valley degree of freedom, which can be used to encode and process quantum information. The valley-selective excitations, governed by the circularly polarised light resonant with the material's band-gap, continues to be the foundation of valleytronics. It is often assumed that achieving valley selective excitation in pristine graphene with all-optical
means is not possible due to the inversion symmetry of the system. Here we demonstrate  that both 
valley-selective excitation and valley-selective high-harmonic generation
can be achieved in pristine graphene by using the combination of two
counter-rotating circularly polarized fields, the fundamental and its second harmonic. 
Controlling the relative phase between the two colours allows us to select the valleys
where the electron-hole pairs  and higher-order harmonics are generated. We also describe 
an all-optical method for measuring valley polarization in graphene with a weak probe pulse.
This work offers a robust recipe to write and read valley-selective electron excitations in 
materials with zero bandgap and zero Berry curvature.
\end{abstract}

\maketitle 


The realisation of atomically-thin monolayer graphene has led to breakthroughs in 
fundamental and applied sciences~\cite{novoselov2004electric, geim2009graphene}.  
Charge carriers in graphene are described by the massless Dirac equation and 
exhibit exceptional transport properties~\cite{neto2009electronic}, making graphene very attractive 
for novel electronics applications.  
One of the most interesting features of graphene and gapped graphene
materials is the electron's extra degree of freedom, the valley pseudospin, associated
with populating the local minima $\mathbf{K}$ and $\mathbf{K}^{\prime}$ in 
the lowest conduction band of the Brillouin zone.  
This extra degree of freedom has the potential 
to encode, process and store quantum information, opening the 
field of valleytronics~\cite{vitale2018valleytronics}. 

The monolayer graphene, as opposed to 
gapped graphene materials,
presents a fundamental challenge for valleytronics:  it has zero bandgap and  
zero Berry curvature. These aspects are generally considered to be a major impediment
for valleytronics. In gapped graphene materials, valley selectivity is achieved
by matching the helicity of a circularly polarized pump pulse, resonant to the bandgap, to the sign of the 
Berry curvature~\cite{schaibley2016valleytronics, mak2012control, jones2013optical, gunlycke2011graphene, xiao2012coupled}. 
Recently demonstrated~\cite{langer2018lightwave} 
sub-cycle manipulation of electron population 
in  $\mathbf{K}$ and $\mathbf{K}^{\prime}$  valleys of 
tungsten diselenide, achieved with the combination of a resonant pump pulse locked to
the oscillations of the THz control pulse, represents a major milestone. Precise sub-cycle 
control over the driving light fields opens new opportunities for valleytronics,
such as those offered by the new concept of a topological 
resonance, discovered and analysed in Refs.~\cite{motlagh2019topological,motlagh2018femtosecond,kelardeh2016attosecond}.
Single-cycle pulses with the controlled phase of carrier oscillations 
under the envelope offer a route to valleytronics in gapped graphene-type materials 
even when such pulses are linearly, not circularly, polarized ~\cite{jimenez2020sub}.
It is also possible to avoid the reliance on resonances in gapped graphene-type materials, 
breaking the symmetry between the $\mathbf{K}$ and $\mathbf{K}^{\prime}$ 
valleys via a light-induced topological phase transition, closing the 
gap in the desired valley~\cite{jimenez2019lightwave}.

Thus, with its zero bandgap, zero Berry curvature, and identical
dispersion near the bottom of the valleys, pristine
graphene appears unsuitable for valleytronics -- a disappointing conclusion in view of its 
exceptional transport properties. We show that
this generally accepted conclusion is not correct, and that the 
preferential population of a desired valley
can be achieved by tailoring the polarization state of the driving light pulse to
the symmetry of the lattice. Our proposal offers an all-optical route to valleytronics in pristine graphene, 
complementing approaches based on  
creating a gap by using a heterostructure of graphene with hexagonal boron nitride~\cite{gorbachev2014detecting, 
yankowitz2012emergence, hunt2013massive, rycerz2007valley}, or by adding strain and/or defect 
engineering~\cite{grujic2014spin, settnes2016graphene, faria2020valley, xiao2007valley, rodriguez2020floquet}.  

While light configuration we use is similar to that used in
Ref.~\cite{jimenez2019lightwave} for finite bandgap materials,
the physical mechanism underlying valley-selective excitation 
in a zero-band-gap, centrosymmetric material such as graphene is quite different.
In gapped materials, valley polarization is achieved by selectively reducing
the effective bandgap in one of the valleys~\cite{jimenez2019lightwave}.
Here, valley polarization is achieved only when the light-driven electrons 
explore the anisotropic region in the Brillouin zone. 

We also show valley selectivity of harmonic generation in graphene, and demonstrate
it with the same field as we use for valley-selective electronic excitation.  
High-harmonic generation (HHG) is a powerful method for probing attosecond electron 
dynamics in systems as diverse as atoms, 
molecules~\cite{lein2007molecular,baker2006probing,smirnova2009high,shafir2012resolving,bruner2016multidimensional,worner2010following} 
and solids ~\cite{kruchinin2018colloquium, ghimire2011observation, schubert2014sub, langer2016lightwave,
lakhotia2020laser, you2017anisotropic, pattanayak2019direct, luu2015extreme, lanin2017mapping, vampa2015all,
silva2018high,  bauer2018high, reimann2018subcycle, garg2016multi,lakhotia2020laser}. 
In solids,  high harmonic spectroscopy was used to probe 
valence electrons~\cite{lakhotia2020laser}, atomic arrangement in solids~\cite{you2017anisotropic, pattanayak2019direct, lakhotia2020laser}, defects~\cite{sivis2017tailored, mrudul2020high}, 
band dispersion~\cite{ luu2015extreme, lanin2017mapping, vampa2015all} and   
quantum phase transitions~\cite{silva2018high,  bauer2018high, silva2019topological}, and
to realise petahertz electronics in solids~\cite{reimann2018subcycle, garg2016multi, lakhotia2020laser}. 
Last but not least, we describe an 
 all-optical method for measuring valley polarization in graphene with a weak probe pulse.

The key idea of our approach is illustrated in Fig. 1, which shows graphene in real (a) and reciprocal (b)
space, together with the structure of the incident electric field (a) and the corresponding 
vector potential (b). The field is made by superimposing two counter-rotating 
circularly polarized colours at the frequencies $\omega$ and $2\omega$. The Lissajous
figure for the total electric field is a trefoil, and its orientation is controlled by the 
relative two-colour phase $\phi$, i.e., the sub-cycle two-colour delay measured in terms of $\omega$.  
In the absence of the field, the two carbon atoms A and B, in real space,
are related by the inversion symmetry. When the field is turned on, this inversion symmetry
is broken: the electric field in panel (a) always points from the atom A to the one of the 
three neighbouring atoms B during the full laser cycle, but not the other way around.
Indeed, if the centre of the Lissajous figure is placed on the atom B, the field points
in the middle between its neighbours.
One can control this symmetry breaking by rotating the trefoil, interchanging the
roles of the atoms A and B. Thus, the bi-circular field offers simple, all-optical, ultrafast 
tool to break the inversion symmetry of the graphene lattice in a controlled
way. Such controlled symmetry breaking  allows one to control the relative 
excitation probabilities induced by the same laser field in the 
adjacent $\mathbf{K}$ and $\mathbf{K}^{\prime}$ valleys of the Brillouin zone. 

\begin{figure}[h!]
\includegraphics[width= 15cm]{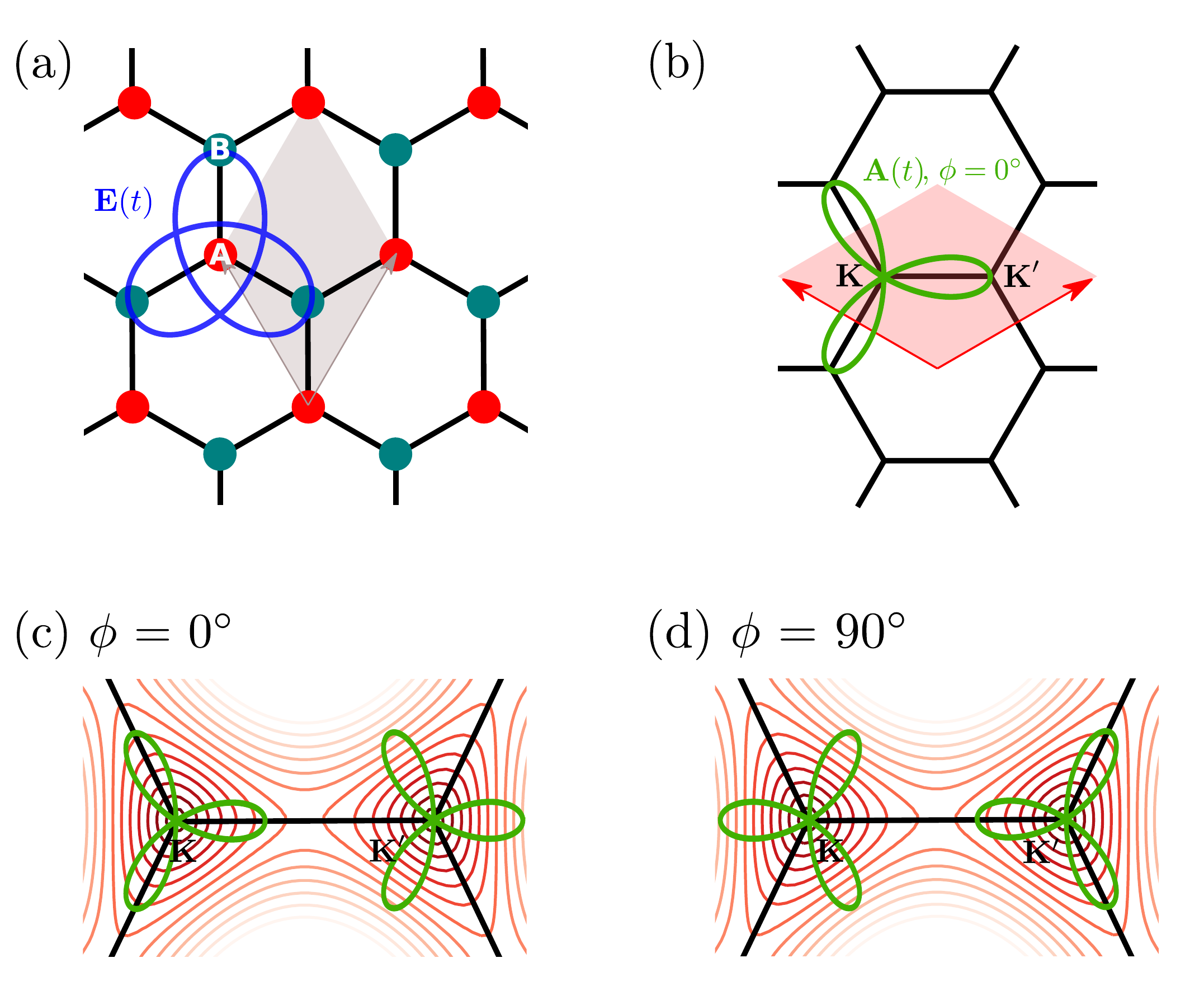}
\caption{{Physical mechanism of valley polarization in graphene.}   
(a) Graphene lattice in the coordinate space, with the Lissajous 
figure of the bicircular electric field breaking the symmetry between otherwise 
identical carbon atoms  A and B; two-colour phase $\phi = 0$.
(b) Optically induced symmetry breaking viewed 
in the momentum space, the vector potential of the 
bicurcular field is shown for $\phi = 0$. (c, d) Close-up images of the 
valleys show their asymmetry, leading to different laser-driven 
dynamics and different excitation rates
once the electron leaves the bottom of the valley. Here, the red contours show the conduction band energy in the reciprocal space.} 
\label{fig1}
\end{figure} 

Panel (b) provides
the complementary reciprocal-space perspective. In the laser field, the electron crystal momentum
follows  ${\bf k}(t)={\bf k}_i+{\bf A}(t)$, where ${\bf k}_i$ is the initial crystal momentum
and ${\bf A}(t)$ is the laser vector potential, shown in (b) for the 
electric field shown in (a). The asymmetry between the two valleys with respect
to the vector potential is immediately visible. 

Panels (c) and (d) provide additional support to this qualitative picture. 
One can observe that the two field-free 
valleys, $\mathbf{K}$ and $\mathbf{K}^{\prime}$, are only identical near their very bottoms. As soon
as one moves away from the bottom, the valleys start to develop 
the trefoil structure, with $\mathbf{K}$ and $\mathbf{K}^{\prime}$ being the mirror images of each other. 
How the symmetry of the vector potential fits into the symmetry of the 
valley away from their bottoms will control the dynamics and the excitation
probability. 

In sufficiently strong fields, excitation happens not only
at the Dirac point where the gap is equal to zero, but also
in its vicinity. For the vector potential in panel (c), the average
gap seen by the electron when following the vector potential from
the Dirac point in the ${\bf K}$ valley, i.e., moving
along the trajectory ${\bf K}+{\bf A}(t)$,  
is less than in the $\mathbf{K}^{\prime}$ valley, i.e., when 
following the trajectory $\mathbf{K}^{\prime}+{\bf A}(t)$. 
In sufficiently strong and low-frequency fields, such that 
the bandgap along the trajectories ${\bf K}+{\bf A}(t)$ and  
$\mathbf{K}^{\prime}+{\bf A}(t)$ quickly exceeds the photon energy, the 
excitation probability should  be higher in panel
(c). For the same reason, rotating the vector potential as shown in
panel (d) should favour population of the $\mathbf{K}^{\prime}$ valley.
Here, ``quickly'' means ``within a fraction of one-third of the 
laser cycle'', which is the relevant time-scale for the bi-circular field. In this
context, lower laser frequencies leading to higher vector potential are 
better suited to meet this requirement.

In the simulations, we used the nearest-neighbour tight-binding approximation to obtain the ground state of graphene with a 
hopping-energy of 2.7 eV~\cite{reich2002tight}. The lattice parameter of graphene is chosen to be 2.46 \AA. 
The resultant band-structure has zero band-gap with linear dispersion near the two points in the Brillouin zone 
known as $\mathbf{K}$ and $\mathbf{K}^{\prime}$ points. 

The density matrix approach was used to follow the electron dynamics in graphene. 
Time-evolution of density matrix element, $\rho^{\textbf{k}}_{mn}$,  was performed using semiconductor 
Bloch equations within the Houston basis $| n, \textbf{k} + \mathcal{A}(t) \rangle $  as~\cite{golde2008high, floss2018ab}
\begin{equation}\label{eq2}
\partial_{t}  \rho _{mn}^{\textbf{k}}  = -i \epsilon_{mn}^{\textbf{k}+\mathcal{A}(t)}\rho_{mn}^{\textbf{k}} -(1-\delta_{mn})\frac{\rho_{mn}^{\textbf{k}}}{T_2}  + i \mathcal{F}(t) \cdot \left[\sum_l \left(\textbf{d}_{ml}^{~\textbf{k}+\mathcal{A}(t)}\rho_{ln}^{\textbf{k}} - \textbf{d}_{ln}^{~\textbf{k}+\mathcal{A}(t)}\rho_{ml}^{\textbf{k}}  \right) \right]. 
\end{equation}
Here,  $\epsilon^{\textbf{k}}_{mn}$ and  \textbf{d}$_{mn}^\textbf{k}$ are, respectively,  
the band-gap energy and the dipole-matrix elements between $m$ and $n$ energy-bands at \textbf{k}. 
Dipole matrix elements were calculated as \textbf{d}$_{mn}^\textbf{k}$ = -i$\left\langle u_m^{\textbf{k}} |\nabla_{\textbf{k}}| u_n^\textbf{k}\right\rangle$, where $u_n^\textbf{k}$ is the periodic part of the Bloch function.
$\mathcal{F}(t)$ and $\mathcal{A}(t)$ are, respectively, the electric field and vector potential of the laser field and 
are related as $\mathcal{F}(t)$ = -$\partial \mathcal{A}(t)/\partial t$. 
A phenomenological term accounting for the decoherence is added with a constant dephasing time $T_2$. 
Conduction band population relaxation was neglected~\cite{hwang2008single}.

The total current was calculated  as
\begin{equation}\label{eq3}
\mathbf{J}(\textbf{k}, t) = \sum_{m,n} \rho_{mn}^{\textbf{k}}(t)~\textbf{p}_{nm}^{~\textbf{k}+\mathcal{A}(t)}. 
\end{equation}
Here, \textbf{p}$_{nm}^\textbf{k}$ are the momentum matrix elements, obtained as 
$\textbf{p}_{nm}^{\textbf{k}} = \left\langle n, \textbf{k} \left| \nabla_{\textbf{k}} \hat{H}_{\textbf{k}} \right| m, \textbf{k} \right\rangle$. 
The off-diagonal elements of momentum and dipole-matrix elements are related as $\textbf{d}^{\textbf{k}}_{mn}$  =  $i\textbf{p}^{\textbf{k}}_{mn}/\epsilon^{\textbf{k}}_{mn}$.

Finally, the  harmonic spectrum was determined from the Fourier-transform of the time-derivative of the total current as
\begin{equation}
\mathcal{I}(\omega) = \left|\mathcal{FT}\left(\frac{d}{d t} \left[\int_{BZ} \mathbf{J}(\textbf{k},t)~d\textbf{k} \right]\right) \right|^2.
\end{equation}
Here, integration is performed over the entire Brillouin zone.

Valley-selective electronic excitation induced by the tailored field is confirmed by our numerical simulations. In the simulations, graphene is exposed to the bicircular field with the 
vector potential 
\begin{equation}\label{eq4}
\mathcal{A}(t) = \frac{A_0 f(t)}{\sqrt{2}} \left(\left[\cos(\omega t + \phi) + \frac{\mathcal{R}}{2} \cos(2\omega t)\right]\hat{\textbf{e}}_x 
+  \left[\sin(\omega t + \phi) - \frac{\mathcal{R}}{2} \sin(2\omega t)\right]\hat{\textbf{e}}_y\right) .
\end{equation}
Here, $A_0=F_{\omega}/\omega$ is the amplitude of the vector potential
for the fundamental field, $F_{\omega}$ is its strength, $f(t)$ is the temporal envelope of the driving field, 
$\phi$ is the sub-cycle phase difference between the two fields, and $\mathcal{R}$ is the ratio
of the electric field strengths for the two fields, leading to $\mathcal{R}/2$ ratio for the 
amplitudes of the vector potentials.  
The amplitude of the fundamental field was varied up to $F_{\omega}$ = 15 MV/cm, leading to 
the maximum fundamental intensity 3$\times$10$^{11}$ W/cm$^2$, with the fundamental wavelength  
$\lambda = 6 \mu$m. This laser intensity is below the damage threshold  for the monolayer graphene~\cite{yoshikawa2017high,heide2018coherent,higuchi2017light}. 
We have also
varied $\mathcal{R}$, using  $\mathcal{R} = 1$ and $\mathcal{R} = 2$.   
A pulse with sin-squared envelope and 145 femtosecond duration (zero-to-zero)
is employed in this work. Our findings 
are valid for a broad range of wavelengths and field intensities. 
To obtain  the total population of the
different valleys, we have integrated the momentum-resolved population  
over the sections shown in Fig. 2(a). 

\begin{figure}[t!]
\includegraphics[width=15cm]{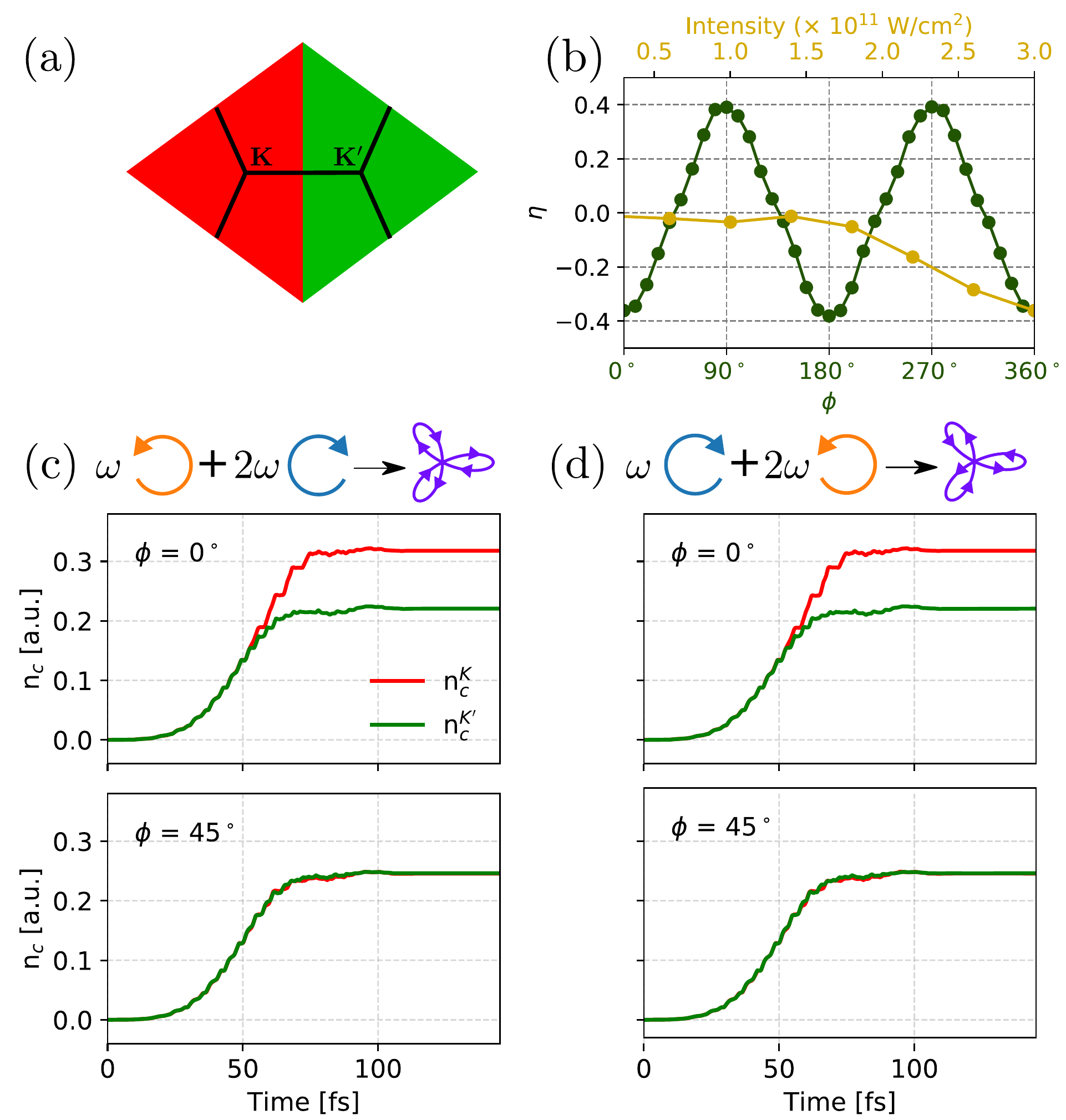}
\caption{{ Valley asymmetry around two valleys  $\mathbf{K}^{\prime}$ and $\mathbf{K}$.} 
(a) Separation of the Brillouin zone into the 
$\mathbf{K}^{\prime}$ and $\mathbf{K}$ valleys. (b) Asymmetry in the 
valley-resolved populations in the conduction band  as a function of $\phi$ (green line with an intensity 3$\times$10$^{11}$ W/cm$^2$), and laser intensity (yellow line with $\phi = 0^{\circ}$) for 
a laser with wavelength of 6 $\mu$m, $\mathcal{R}$ = 2, and a dephasing time  of 10 $fs$. 
(c, d) Excitation dynamics during the laser pulse for 
$\phi  = 0^{\circ}, 45^{\circ}$: switching the
helicities of both the fields simultaneously does not change the outcome.} 
\label{fig2}
\end{figure}

To quantify the amount of the valley polarisation, we used the valley asymmetry  parameter defined 
as 
\begin{equation}\label{eq1}
\eta = \frac{n_{c}^{\mathbf{K}^{\prime}} - n_{c}^{\mathbf{K}}}{(n_{c}^{\mathbf{K}^{\prime}} + n_{c}^{\mathbf{K}})/2}, 
\end{equation}
where $n_{c}^{\mathbf{K}^{\prime}}$ and $n_{c}^{\mathbf{K}}$ are electron populations at the end of the 
laser pulse in the  conduction band 
around   $\mathbf{K}^{\prime}$ and $\mathbf{K}$ valleys, respectively. 

Fig. 2(b) shows the asymmetry in the populations of the ${\bf K}$ and $\mathbf{K}^{\prime}$ valleys as a function
of the two-colour phase $\phi$, for several values of the fundamental field
amplitude and $\mathcal{R} = 2$. Substantial contrast between the two valleys is achieved
once the excitations leave the bottom of the Dirac cone, with values as high
as $\pm 36 \%$ for $\phi  = 0^{\circ}, 90^{\circ}$, and
no asymmetry at  $\phi  = 45^{\circ}$. Here, each 180$^\circ$ change in $\phi$ results 
in 120$^\circ$ rotation of the trefoil, yielding an equivalent configuration. 
This is the reason for the periodicity of the valley asymmetry presented in Fig.~\ref{fig2}(b).
The higher-populated valley is the one where the  vector potential ``fits'' better into the 
shape of the valley, minimizing the bandgap along the electron trajectory
in the momentum space. The same results are obtained when simultaneously 
changing the  helicities of both driving fields, Fig. 2(c, d).
The asymmetry in the valley population is negligible up to an 
intensity of $2 \times 10^{11}$ W/cm$^2$ 
and gradually increases with intensity [Fig. 2 (b)]. 
This shows that the valley asymmetry is observed only when the laser pulse is able 
to drive the electrons to the anisotropic part of the conduction band. 

The current generated by the electron injection into the conduction band valleys is 
accompanied by harmonic radiation 
and makes substantial contribution to the lower order harmonics, 
such as H4 and H5 for our two-colour driver. These harmonics are stronger 
whenever the dispersion $\epsilon({\bf k})$ is
more nonlinear. In this respect, for the electrons following the trajectories ${\bf K}+{\bf A}(t)$ and  
$\mathbf{K}^{\prime}+{\bf A}(t)$, the current-driven high harmonic generation from 
the  $\mathbf{K}^{\prime}$ valley 
is preferred for the vector potential in Fig. 1(c).  Conversely, for the 
vector potential in Fig. 1(d) low-order, current-driven 
harmonic generation should be preferred from the ${\bf K}$ valley. Indeed, 
following the vector potential, the
electron is driven against the steeper walls in the $\mathbf{K}^{\prime}$ valley in panel (c) and 
against the steeper walls in the ${\bf K}$ valley in panel (d). 

\begin{figure}[h!]
\includegraphics[width= 18 cm]{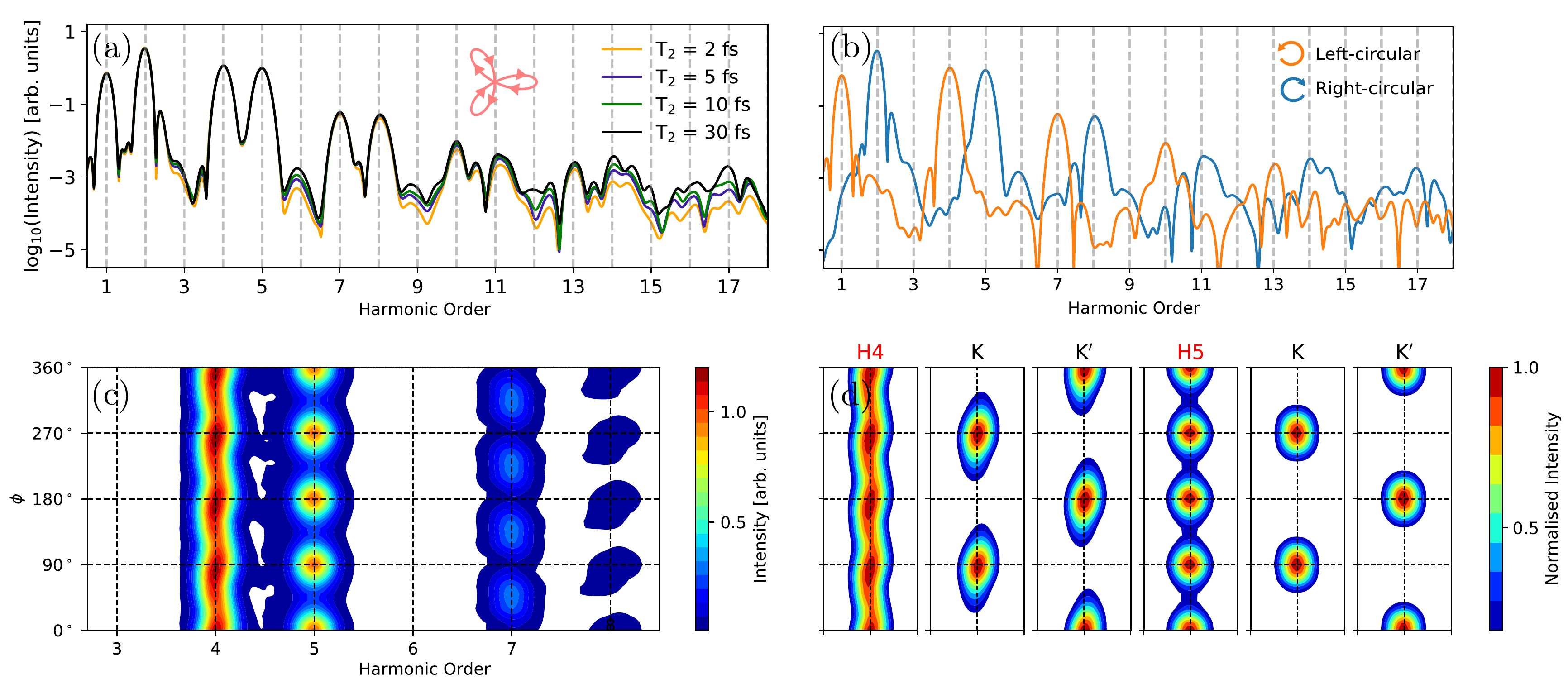}
\caption{{Valley polarization in high-harmonic emission driven by a bicircular field in graphene.}   
(a) Harmonic spectrum for different dephasing times, for the 
vector potential shown in the inset ($\phi = 0$);
(b) Polarisation-resolved high harmonic emission, 
for left-handed circularly polarised 
fundamental  (orange) and right-handed circularly polarised 
second harmonic (blue). Alternating harmonics follow alternating 
helicities: $3n+1$ follow the fundamental,  
$3n+2$ follow the second harmonic,  the $3n$ harmonics 
missing due to symmetry. 
(c) Harmonic emission as a function of the 
two-colour phase $\phi$. (d) Valley polarization of H4 and H5 as a function
of the two-colour phase.
The total field is identical for $\phi = 0^{\circ}$ and $\phi = 180^{\circ}$ owing to the threefold symmetry.} 
\label{fig3}
\end{figure} 

These qualitative expectations are also confirmed by our 
numerical simulations, shown in Fig. 3.
The same laser parameters used in Fig.~\ref{fig2} are used in this simulation.

In general, the interband and intraband harmonic emission mechanisms in graphene are 
coupled~\cite{taucer2017nonperturbative,liu2018driving,al2014high}, except 
at low electric fields~\cite{al2014high}, leading to a complex interplay between the 
interband and intraband emission mechanisms.  The former should be
more sensitive to the dephasing than the latter. To this end, we have calculated the 
the harmonic spectrum for different dephasing 
times $T_2$, see Fig. 3(a). We find that the harmonic emission is 
essentially $T_2$-independent, until at least H13, suggesting that
the intraband emission mechanism is generally dominant.

Fig. 3(b) shows polarisation-resolved high-harmonic spectrum. The
$(3n+1)$ harmonics follow the polarisation of the $\omega$ pulse (left-handed circular polarisation), whereas 
$(3n+2)$ harmonics follow the polarisation of the $2 \omega$ pulse (right-handed circular polarisation), while  
$3n$ harmonics are missing, just like in atomic media~\cite{fleischer2014spin,neufeld2019floquet}.
In this context, we note that while harmonic generation with
single-colour circularly polarized drivers is forbidden in atoms, 
such selection rules do not generally arise in solids~\cite{ghimire2011observation}. 
However, the ellipticity-dependence 
studies on graphene show very weak harmonic yield for drivers with higher ellipticity~\cite{yoshikawa2017high, taucer2017nonperturbative, liu2018driving}, as we have  
also observed. As in atoms, application of bicircular fields allows for efficient 
generation of circularly polarized harmonics in graphene.

Fig. 3(c) shows the dependence of harmonic generation on the orientation of
the vector potential relative to the structure of the Brillouin zone.
As expected, the total harmonic yield is modulated as the trefoil is
rotated, with the lower-order current-driven harmonics H4 and H5 following the expected
pattern, maximizing when the electrons are driven into the steeper walls
in either the $\mathbf{K}^{\prime}$ or ${\bf K}$ valley. 

The contribution of different valleys to H4 and H5 as a function of
the field orientation is presented in Figs. 3(d). Consistent with the qualitative analysis above, 
maximum harmonic contribution of the $\mathbf{K}^{\prime}$
valley corresponds to the vector potential orientations such as shown in Fig. 1(c),
while the maximum contribution of the ${\bf K}$
valley corresponds to the vector potential orientations such as shown in Fig. 1(d). Therefore, we are able to control the valley-polarisation of 
the harmonics by controlling the two-colour phase $\phi$. We have also checked
that these results do not depend on the specific directions of rotation of the 
two driving fields. That is, we find the same results when simultaneously changing 
the helicities of both driving fields. 

\begin{figure}[t!]
		\includegraphics[width= 15 cm]{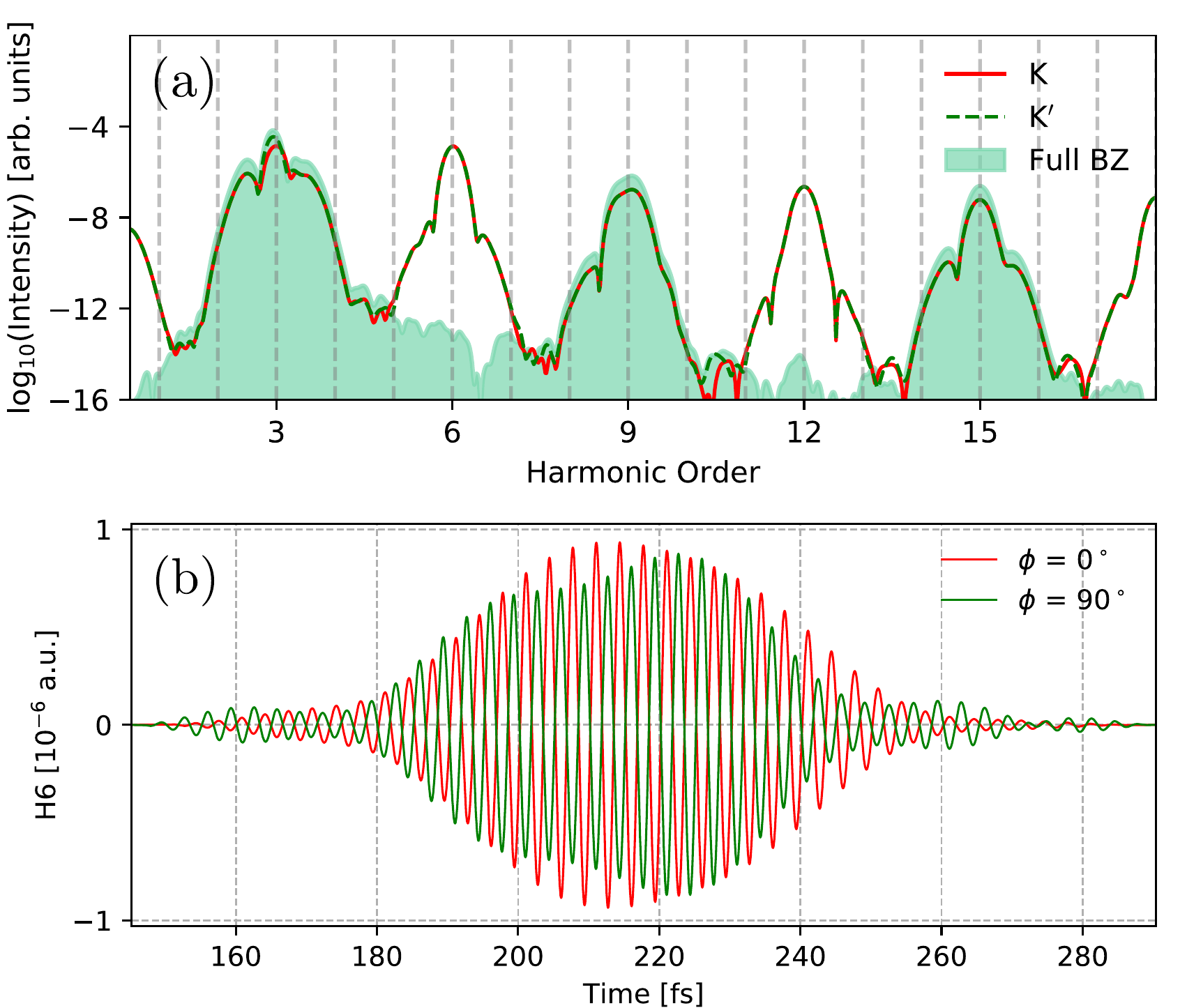}
	\caption{{ Read-out of the induced valley-polarization in graphene.}
		(a) Harmonic spectrum corresponding to a linearly polarised field with an amplitude of 1.5 MV/cm and a frequency of 3$\omega$ for monolayer graphene. The red and green lines show the valley-resolved HHG spectrum.  (b) The 6$^{\textrm{th}}$ harmonic (H6) generated by 3$\omega$  pulse after the bicircular field broke the inversion symmetry in graphene.}
	\label{fig4}
\end{figure}

To read-out the induced valley polarization,  we employ a probe 
pulse of  frequency 3$\omega$  linearly polarized  
along the x direction (parallel to $\Gamma$-$\mathbf{K}$ and $\mathbf{K'}$-$\mathbf{K}$  
directions in the Brillouin zone). 
The amplitude of the probe field is 1.5 MV/cm (F$_{\omega}$/10). 
Since the $\mathbf{K', K}$ valleys in graphene are related by space inversion,  
the even-order harmonics generated by individual (asymmetric) valleys are equal in magnitude 
but opposite in phase, see Fig.~\ref{fig4}(a) (red and green lines). 
In the absence of valley polarization, their interference leads to
complete cancellation of even harmonics, Fig.~\ref{fig4}(a) (full BZ signal). 
In the presence of valley polarization, the symmetry is broken, the 
cancellation of even harmonics is quenched, and even harmonics 
signal scales proportional to valley polarization (see also~\cite{golub2014valley} and ~\cite{jimenez2020sub}). 
The phase of the even harmonics follows 
the dominant valley. Importantly, 3$N\omega$ harmonics are 
absent  in the spectra generated by the bicircular $\omega-2\omega$ field, Fig.~\ref{fig3}(b).
Thus, even harmonics generated by the 3$\omega$ probe 
provide background-free measurement of valley polarization. 

Figure ~\ref{fig4}(b) shows generation of the second harmonic of the 
3$\omega$ probe pulse (labelled H6) for the two-color phases of the bi-circular pump
$\phi$ = 0$^\circ$ (red curve) and  90$^\circ$ (green curve), which switches the 
valley polarization  between  $\mathbf{K}$ and $\mathbf{K'}$  valleys. While the 
H6 intensity measures valley polarization, its phase clearly identifies the dominant valley. This phase 
can be measured by interfering the signal with the reference second harmonic of 3$\omega$
generated, e.g., from a BBO crystal. Controlling the delay of the 
reference second harmonic generated in the BBO crystal,  we can 
map the phase of H6 generated by graphene
on the amplitude modulation of their interference.

In summary, valley polarisation in both electronic excitation and harmonic generation
can be achieved in pristine graphene by 
tailoring the Lissajous figure of the driving pulse to the symmetry of the graphene
lattice. This allows one to both break the inversion symmetry between the adjacent carbon atoms and also 
exploit the anisotropic regions in the valleys, taking advantage of
the fact that the energy landscape of the valleys are mirror images of each other.  
Present work opens an avenue for a new regime of valleytronics in pristine graphene and similar materials with zero 
bandgap and zero Berry curvature.

\section*{Funding}
M.I. acknowledges support from the Deutsche Forschungsgemeinschaft (DFG) Quantum Dynamics in Tailored Intense Fields (QUTIF) grant.
M.I. and A. J-G acknowledge funding from the European Union's Horizon 2020 research and innovation programme 
under grant agreement No 899794, ``Optologic.''
G. D. acknowledges support from Science and Engineering Research Board (SERB) India 
(Project No. ECR/2017/001460).

\section*{Disclosures}
The authors declare no conflicts of interest.


\end{document}